\documentclass[aps,amsmath,amssymb,reprint, double column,prl, showkeys]{revtex4-2}

\usepackage{graphicx,epsfig}
\usepackage{amssymb}
\usepackage{amsmath}
\usepackage{bm}
\usepackage{textcomp}
\usepackage{color}
\usepackage{float}
\usepackage{comment}

\usepackage{soul}

\begin{document}
\setcounter{page}{1}

\title[]{Delocalization and Universality of the Fractional Quantum Hall Plateau-to-Plateau Transitions}
\author{P. T. \surname{Madathil}}
\author{K. A. \surname{Villegas Rosales}}
\author{C. T. \surname{Tai}}
\author{Y. J. \surname{Chung}}
\author{L. N. \surname{Pfeiffer}}
\author{K. W. \surname{West}}
\author{K. W. \surname{Baldwin}}
\author{M. \surname{Shayegan}}
\affiliation{Department of Electrical Engineering, Princeton University, Princeton, New Jersey 08544, USA}

\date{\today}

\begin{abstract}

Disorder and electron-electron interaction play essential roles in the physics of electron systems in condensed matter. In two-dimensional, quantum Hall systems, extensive studies of disorder-induced localization have led to the emergence of a scaling picture with a single extended state, characterized by a power-law divergence of the localization length in the zero-temperature limit. Experimentally, scaling has been investigated via measuring the temperature dependence of plateau-to-plateau transitions between the integer quantum Hall states (IQHSs), yielding a critical exponent $\kappa\simeq 0.42$. Here we report scaling measurements in the fractional quantum Hall state (FQHS) regime where interaction plays a dominant role. Our study is partly motivated by recent calculations, based on the composite fermion theory, that suggest identical critical exponents in both IQHS and FQHS cases to the extent that the interaction between composite fermions is negligible. The samples used in our experiments are two-dimensional electron systems confined to GaAs quantum wells of exceptionally high quality. We find that $\kappa$ varies for transitions between different FQHSs observed on the flanks of Landau level filling factor $\nu=1/2$, and has a value close to that reported for the IQHS transitions only for a limited number of transitions between high-order FQHSs with intermediate strength. We discuss possible origins of the non-universal $\kappa$ observed in our experiments.

\end{abstract}

\maketitle  

In 1958, P. W. Anderson introduced the theory of localization in disordered systems \cite{anderson1958absence}. He showed that in sufficiently dilute systems with only short-range forces, states return to their original site with a finite probability in the long-time limit and thus, there is an absence of diffusion. While the scaling theory of localization predicts the lack of extended states in two dimensions \cite{abrahams1979scaling}, quantum Hall systems are reported to host both localized and extended states \cite{aoki1985critical,chalker1988percolation,wei1986localization,goldman1990nature}. In the zero-temperature limit, as the Fermi energy approaches a single critical energy ($E_c$), theory predicts that the localization length ($\xi$) diverges according to the power law $\xi \propto |E - E_c|^{-\gamma}$ with a universal critical exponent $\gamma$ \cite{pruisken1988universal,huo1993universal,huckestein1995scaling}. Criticality is also associated with fundamental phenomena such as anamolous diffusion, multifractal conductance fluctuations, and power-law-density-correlations \cite{chalker1988scaling,pook1991multifractality,amin2022multifractal} owing to the large fluctuations in the local densities and currents in the absence of a length scale. Since its inception, the theory of criticality for the non-interacting integer quantum Hall states (IQHSs) has garnered immense interest, with recent numerical calculations suggesting substantial corrections to the critical exponent and predicting model-dependent exponents \cite{abrahams1979scaling,dresselhaus2021numerical,dresselhaus2022scaling,zhu2019localization,zirnbauer2019integer}. 

The strongly-interacting nature of the \textit{fractional} quantum Hall states (FQHSs) poses a more challenging theoretical framework in understanding critical phenomena. Exact-diagonalization studies are limited to very small systems and are often inadequate in capturing the dynamics in the thermodynamic limit. The composite-fermion (CF) theory provides a fruitful way to distill the physics of the FQHSs by treating the system of strongly-interacting electrons as a collection of weakly-interacting, magnetic-flux-electron quasi-particles, namely the CFs \cite{jain1989composite,jain2007composite}. The simplest FQHSs occur in the lowest
Landau level, flanking the filling factor $\nu$ = 1/2
at $\nu = \frac{p}{2p \pm 1}$ where $p$ is a positive integer. The FQHS at a particular $\nu$ can then be thought of as the $p^{th}$ IQHS of CFs. \cite{jain1989composite,jain2007composite}. Early theoretical work suggested the same scaling 
 exponents for the transitions between FQHSs as those for the IQHSs \cite{jain1990scaling} but a microscopic confirmation of this correspondence was lacking. More recent, rigorous calculations elaborate on the correspondence and highlight similar localization physics in the two regimes \cite{pu2022anderson,hui2019non,kumar2022interaction}.

\begin{figure*}[t]
\centering
\includegraphics[width=1\textwidth]{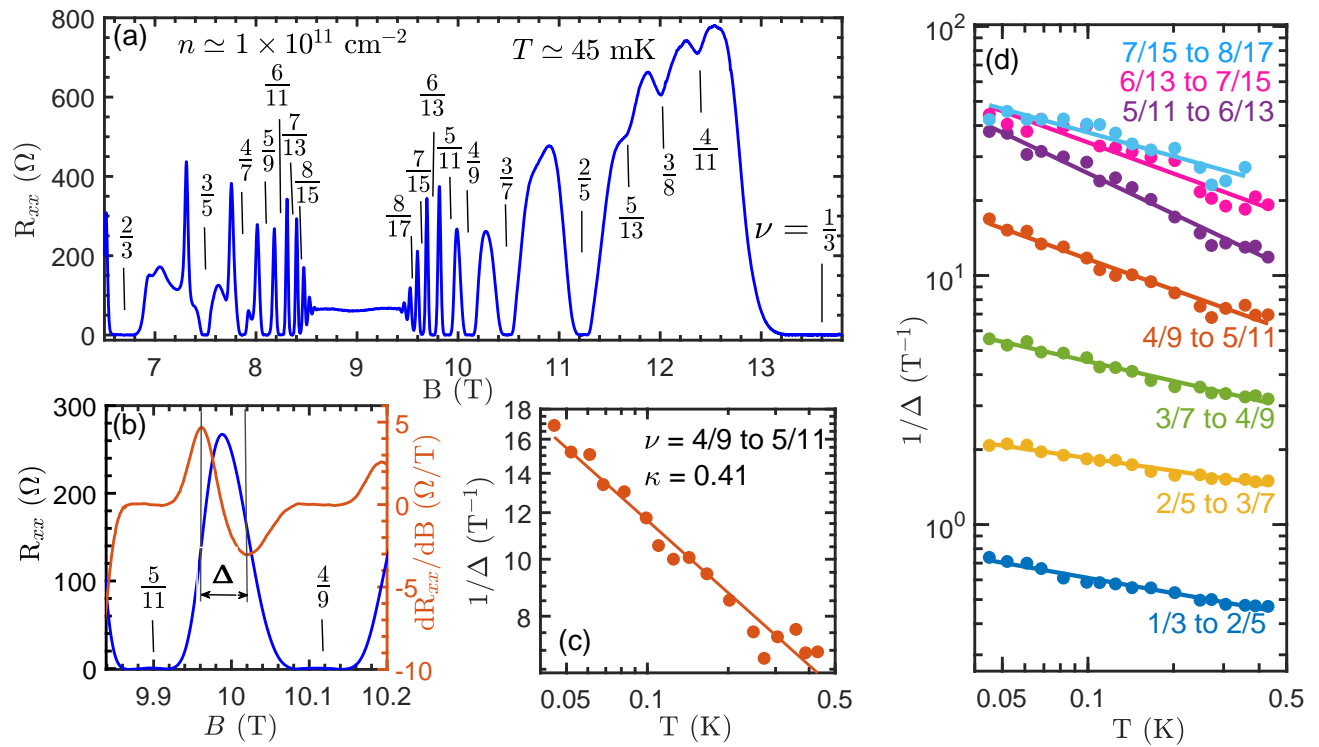}
\centering
  \caption{\label{Ip} 
(a) Longitudinal resistance ($R_{xx}$) vs. magnetic field ($B$) for a 2DES confined to a 30-nm-wide GaAs QW at $T\simeq$ 45 mK. (b) The blue trace is the smoothed $R_{xx}$ between $\nu = 4/9$ and 5/11. The red trace is the corresponding $dR_{xx}/dB$ vs. $B$. The vertical grey lines mark $B$ at which $dR_{xx}/dB$ has an extremum, and the field difference between the two extrema is denoted by $\Delta$. (c)  Log-log plot of $1/\Delta$ vs. $T$ for the transition between the $\nu = 4/9$ and $5/11$ FQHSs; the red line is a least-squares-fit through the data points according to $1/\Delta \propto T^{-\kappa}$ and $\kappa$ is the magnitude of the slope extracted from the fit. (d) Log-log plot of $1/\Delta$ vs. $T$ for the 30-nm-wide QW for the different FQHS transitions; the lines are fits to the data points. 
  }
  \label{fig:Ip}
\end{figure*}

Experimentally, one can measure the divergence of $\xi$ via studying the temperature (\textit{T}) dependence of the Hall ($R_{xy}$) and longitudinal ($R_{xx}$) resistances at the transitions between the QHS plateaus. The derivative of $R_{xy}$ (with respect to the magnetic field, \textit{B}) at the critical magnetic field, and the inverse of the half-width of $R_{xx}$ between two successive quantum Hall states $(1/\Delta)$, both diverge according to the power law $T^{-\kappa}$. The quantum phase coherence length ($L_\phi$) also diverges with temperature as $L_\phi \propto T^{-q/2}$, and the three exponents $\kappa$, $q$ and $\gamma$ follow the relation $\kappa = \frac{q}{2\gamma}$ \cite{pruisken1988universal,huckestein1995scaling,wei1988experiments,wei1994current,koch1991size,li2005scaling,li2009scaling}. Despite some discrepancies in earlier studies, systematic measurements for the transitions between the IQHSs have concluded a value of $\kappa \simeq 0.42$, in excellent agreement with theoretical expectations \cite{li2005scaling,li2009scaling}. Experimental scaling measurements in the FQHS regime, however, are quite scarce. An early study on a sample with relatively low mobility suggested that the transition between the strongest FQHSs (at $\nu=1/3$ and 2/5) has the same exponent of criticality ($\kappa$) as the transitions between the IQHSs \cite{engel1990critical}, but a complete set of exponents to test universality for transitions between various, high-order FQHSs is still lacking. The focus of this Letter is to investigate scaling in ultra-high-quality GaAs two-dimensional electron systems (2DESs) in the FQHS regime.

Our experiments were performed on a series of 2DESs confined to GaAs quantum wells (QWs) of well widths 30 to 50 nm with densities $\simeq$ $1 \times 10^{11}$ cm$^{-2}$ \cite{chung2021ultra,chung2022understanding}. This was achieved by flanking the QWs with 220-nm-thick Al$_{0.24}$Ga$_{0.76}$As barriers and placing the Si doping layers inside doping wells \cite{chung2020working}. The mobilities in these samples are $ \simeq 20 \times 10^6$ cm$^2$V$^{-1}$s$^{-1} $. The samples were then cooled in a dilution refrigerator and magnetoresistance measurements were carried out using standard lock-in techniques. The samples had a van der Pauw geometry, with alloyed InSn contacts at the corners and midpoints of edges of $4 \times 4$ mm$^2$ square pieces.

In Fig. 1(a), we show the $R_{xx}$ vs. $B$ trace for the 30-nm-wide GaAs QW at $T \simeq$ 45 mK. The exceptional sample quality is seen from the presence of a series of FQHSs at $\nu = \frac{p}{2p \pm 1}$ around $\nu=1/2$ extending up to \textit{p} = 10, namely $\nu=10/21$  on the electron side ($\nu<1/2)$ and $\nu=10/19$ on the hole side ($\nu>1/2)$. We observe well-developed FQHSs, with vanishingly small $R_{xx}$, for states from $\nu=1/3$ to $6/13$ and $\nu=2/3$ to $6/11$. We also see emerging $R_{xx}$ minima between $\nu=1/3$ and $2/5$ at $\nu=4/11$, $3/8$, and $5/13$ which correspond to the FQHSs of CFs in an interacting CF picture \cite{wojs2007spin,mukherjee2014possible,pan2015fractional,samkharadze2015observation}.

Figures 1(b) and (c) describe the procedure employed in extracting the critical exponent, $\kappa$, from the dependence of $R_{xx}$ on $B$. The blue trace in Fig. 1(b) shows $R_{xx}$ vs. $B$ between $\nu=5/11$ and $4/9$. We first employ a Savitzky-Golay filter \cite{savitzky1964smoothing} with order 2 to smooth out the raw data shown in Fig. 1(a). We then determine $dR_{xx}/dB$, as shown in red. The extrema in $dR_{xx}/dB$, corresponding to the highest rate of change in resistance with \textit{B} between the two FQHSs, are marked by the two vertical grey lines. The difference between the magnetic fields at which $dR_{xx}/dB$ has an extremum is defined as $\Delta$. We repeat this procedure for a range of temperatures and proceed to extract $\kappa$ as shown in Fig. 1(c). The circles correspond to $1/\Delta$ obtained at different \textit{T} and are shown in a log-log plot. The line is a least-squares-fit to the data points and the magnitude of its slope yields $\kappa$. We then proceed to analyze the temperature dependence of $1/\Delta$ for the transitions between different FQHSs, as shown in Fig. 1(d). While $1/\Delta$ for all the transitions exhibit linear dependencies on $T$ in log-log plots, the slopes and thus $\kappa$ are strikingly different.

 \begin{figure*}[t]
\centering
\includegraphics[width=1\textwidth]{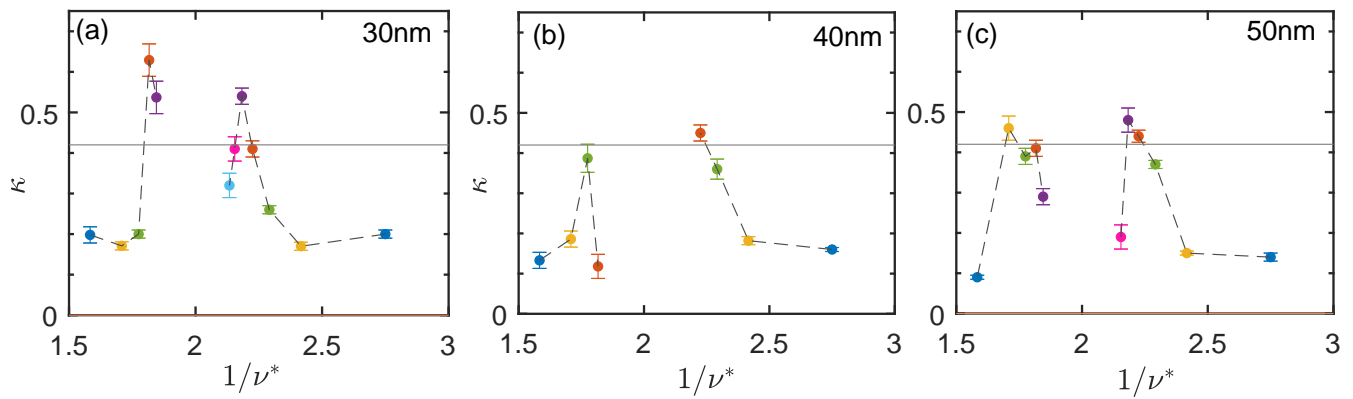}
\centering
  \caption{\label{Ip} 
(a) The extracted $\kappa$ for the 30-nm-wide QW are plotted vs. $1/\nu^*$, defined as $1/\nu^* = (1/\nu_1 + 1/\nu_2)/2$, where $\nu_1$ and $\nu_2$ are the fillings of two consecutive FQHSs, e.g., $\nu_1$ = 1/3 and $\nu_2$ = 2/5 yield a value of 1/$\nu^*$ = 2.75. The colors of data points for $2<1/\nu^*<3$ represent the colors of data presented in Fig. 1(d) for different transitions. The dashed lines connecting the data points are guides to the eye. The grey, horizontal line at $\kappa = 0.42$ represents the expected exponent. (b) and (c) summarize the extracted $\kappa$ vs. 1/$\nu^*$ for the 40- and 50-nm-wide QWs.
  }
  \label{fig:Ip}
\end{figure*}
 
 A summary of all the extracted $\kappa$ vs. 1/$\nu^*$ is shown in Fig. 2(a) for the 30-nm-wide QW sample; similar data for the 40- and 50-nm-wide samples are shown in Figs. 2(b,c). For the x-axis of Figs. 2(a,b,c), we use the harmonic mean of the filling factors ($\nu_1$ and $\nu_2$) of two successive FQHSs, namely $1/\nu^* = (1/\nu_1 + 1/\nu_2)/2$. The grey horizontal lines indicate $\kappa$ = 0.42, expected from measurements and calculations for IQHSs. Our experimentally-extracted $\kappa$ for the FQHS transitions, however, exhibit a non-universal and non-monotonic behavior. For the transitions between the strongest FQHSs (farthest away from $\nu = 1/2$), $\kappa$ is much smaller than 0.42. As we move towards $\nu = 1/2$, $\kappa$ increases dramatically and reaches maximum values that exceed 0.42. It then decreases again as $\nu$ approaches 1/2. The trend for the evolution of $\kappa$ on the hole side ($\nu > 1/2$) is qualitatively similar to its electron counterpart ($\nu < 1/2$).

The exponent $\kappa$ can also be extracted from the temperature dependence of the Hall resistance ($R_{xy}$). The maximum value of the derivative of $R_{xy}$ with respect to $B$, at the critical magnetic field ($B_c$), which corresponds to the critical energy, exhibits a power-law divergence with temperature, with the same critical exponent, i.e., $\frac{dR_{xy}}{dB}|_{B = B_c} \propto T^{-\kappa}$ \cite{pruisken1988universal,huckestein1995scaling,wei1988experiments,li2005scaling}. We report the values of $\kappa$ extracted from $R_{xy}$ for the 50-nm-wide QW in the Supplemental Material \cite{supplementary} and show that they closely follow the values obtained from $R_{xx}$.

In order to discuss Fig. 2 data, we first briefly review what is known for the localization and scaling in the IQHS and FQHS regimes. For the IQHS case, numerous theoretical attempts have been made to determine the value of the critical exponent $\gamma$ that quantifies the divergence of the localization length at transitions between the plateaus \cite{pruisken1988universal,huo1993universal,huckestein1995scaling, zirnbauer2019integer, zhu2019localization, dresselhaus2021numerical, dresselhaus2022scaling,Arapov2019review}. While different models of localization predict slightly dissimilar values for $\gamma$, it is generally found that $\gamma \simeq 2.4$ \cite{huckestein1995scaling}. Assuming a value of $q \simeq 2$ for the exponent of the phase coherence length, $\gamma \simeq 2.4$ implies that $\kappa \simeq 0.42$. Experimentally, early studies on 2DESs confined to different materials (Si-MOSFETs, In$_x$Ga$_{1-x}$As, and GaAs) provided different values for $\kappa$, deduced from the $T$-dependence of the plateau-to-plateau transition widths \cite{wakabayashi1990experiments,wei1988experiments,koch1991experiments,koch1991size,d1992full,hwang1993scaling,wei1992effect,gusev1992percolation,Saeed2014frequency,Dodoo2014nonuniversality}. While in some specific materials and for certain transitions, a $\kappa \simeq 0.42$ was indeed measured, this was not found to be universal; see Ref. \cite{huckestein1995scaling} for a comprehensive review of early results. Later systematic studies by Li \textit{et al.} \cite{li2005scaling, li2009scaling}, performed on 2DESs confined to Al$_y$Ga$_{1-y}$As samples shed new light on the experimental situation. They demonstrated that for these samples, where the dominant electron scattering mechanism is the short-range, alloy scattering, the scaling exponents are indeed universal and have values $\kappa \simeq 0.42$, $q \simeq2$, and $\gamma \simeq 2.4$, very much consistent with the theoretical expectations.

Considerably less in known for the transitions between the plateaus in the FQHS case. An early experimental study by Engel \textit{et al.} \cite{engel1990critical} reported $\kappa \simeq 0.43$ for the transition between the $\nu=1/3$ and $2/5$ FQHSs, i.e., a value very close to the IQHS case. Note that the density of the sample used in Ref. \cite{engel1990critical} was very close to the density of our sample, but the quality was much inferior as judged by its much (about 20 times) lower mobility and the presence of only very few FQHSs, namely those at $\nu=1/3$, 2/5, 2/3, and 3/5. Very recently, the transitions between FQHSs were studied theoretically by Pu \textit{et al.} \cite{pu2022anderson} in a non-interacting CF formalism, and it was concluded that the critical exponents for these transitions should be the same as in the IQHS regime, confirming the data of Ref. \cite{engel1990critical}. Note that the conclusion of Ref. \cite{pu2022anderson} can be readily understood: In a non-interacting CF picture, the FQHSs can be simply mapped into the IQHSs of CFs.

Now our data in Fig. 2 reveal that $\kappa \simeq 0.20$ for the $\nu$ = 1/3 to 2/5 transition (1/$\nu^*$ = 2.75). This is significantly smaller than the theoretically-expected value of 0.42 \cite{pu2022anderson}, or previously reported in experiments of Ref. \cite{engel1990critical} ($\simeq 0.43$). The discrepancy likely stems from the much higher quality of our present samples and the fact that they exhibit numerous developing FQHSs between $\nu=1/3$ and 2/5 [see Fig. 1(a)]. These FQHSs at intermediate fillings between the standard, Jain-series FQHSs (i.e., those at $\nu = 1/3$ and 2/5) are a common feature of ultra-high-quality samples such as ours, and can be described as the FQHSs of CFs, originating from interaction between CFs \cite{wojs2007spin,mukherjee2014possible,pan2015fractional,samkharadze2015observation}. Note that such additional FQHSs are completely absent in the sample of Ref. \cite{engel1990critical} which exhibits only a single, sharp maximum in $R_{xx}$ between the deep and wide $R_{xx}$ minima at $\nu=1/3$ and 2/5. As a result, $\Delta$ is significantly smaller in Ref. \cite{engel1990critical} and, more importantly, $1/\Delta$ diverges faster as temperature approaches zero. In contrast, in our much better quality 2DESs, the growth of $1/\Delta$ at low $T$ is limited by the presence of these intermediate FQHSs, rendering $\kappa \simeq 0.2$; see also Supplemental Material \cite{supplementary}. It is worth mentioning that, in some of the experiments in the IQHS regime on samples where the spin splitting in Landau Levels was not well resolved, a $\kappa \simeq 0.21$ was also found for transitions between two IQHSs which were separated by a weakly-developed or undeveloped IQHS \cite{hwang1993scaling}. 

In our sample, as seen in Fig. 1(a), emerging features are also seen at transitions between other consecutive Jain-series FQHSs: $\nu=2/5$ to 3/7, 2/3 to 3/5, 3/5 to 4/7, and 4/7 to 5/9. The measured $\kappa$ for these transitions are also $\simeq 0.2$, much smaller than 0.42 [Fig. 2(a)], consistent with our conjecture that the presence of intermediate features in the transition region is the cause of smaller than expected $\kappa$.

In Fig. 2(a) we also observe a decrease of $\kappa$ for the transitions between the highest-order FQHSs closest to $\nu=1/2$, e.g., between 7/15 and 8/17. While we do not know the reason for this decrease, it is worth noting that these FQHSs are not well-developed even at the lowest temperatures achieved in our experiments. They are akin to the weak, high-filling-factor IQHSs, more appropriately termed Shubnikov-de Haas oscillations, seen near zero magnetic field. The apparent decrease we observe in $\kappa$ as $\nu=1/2$ is approached might be related to this weakness of the highest-order FQHSs.

The non-universality of $\kappa$ we measure and its deviations from the expected value might also be related to the type of disorder present in our samples. Experiments in the IQHS regime have indeed shown that the nature of the disorder in the 2DES does play an important role in determining the value of $\kappa$ and its universality. Li \textit{et al.} \cite{li2005scaling,li2009scaling} performed a systematic localization study in 2DESs confined to Al$_y$Ga$_{1-y}$As alloy QWs (rather than single-crystal GaAs QWs) with different Al alloy compositions $y$. Their results revealed that the scaling follows the theoretical power-law only in the range $0.0065 \leq y \leq 0.016$ where the disorder and electron scattering are dominated by short-range alloy potential fluctuations. In contrast to their samples, the primary contributions to disorder in the ultra-high-quality 2DESs studied in our experiments come from remote and background (residual) ionized impurities \cite{chung2021ultra,chung2022understanding}. These lead to long-range potential fluctuations. Indeed, in GaAs 2DESs similar to ours, with long-range disorder, Wei \textit{et al.} \cite{wei1992effect} reported significant deviations from the theoretically-expected $\kappa$ in the IQHS regime. While it is in principle possible to fabricate 2DESs confined to Al$_y$Ga$_{1-y}$As QWs and study localization phenomena in the FQHS regime, the experiments would be challenging: $y$ has to be sufficiently large to induced significant alloy disorder, and yet small enough to preserve the quality of the 2DES at low densities so that FQHSs could be still observed at accessible magnetic fields \cite{footnote}. %(Samples of Ref. \cite{li2005scaling,li2009scaling}, e.g., had very large 2DES densities and the FQHS regime near $\nu=1/2$ could not be reached at reasonable magnetic fields.)

In summary, we report values of the critical exponent $\kappa$ for transitions between the plateaus of FQHSs flanking $\nu=1/2$ in ultra-high-quality GaAs 2DES samples. Several samples with different QW widths exhibit a qualitatively similar behavior: $\kappa$ changes non-monotonically as a function of filling and, only for a limited number of transitions between high-order FQHSs with intermediate strength, has a value close to $\simeq$ 0.42, the value predicted theoretically based on a non-interacting CF picture. The non-universality of $\kappa$ might be a result of the additional, unconventional FQHSs that emerge between the neighboring, strong, Jain-sequence FQHSs when CFs are interacting. It can also be a consequence of the nature of the disorder in the samples. Our results shed light on the complex role of interaction, and highlight the need for future experimental and theoretical efforts to understand the physics of criticality for the FQHS plateau-to-plateau transitions.

\begin{acknowledgments}

We acknowledge support by the National Science Foundation (NSF) Grant No. DMR 2104771 for measurements. For sample characterization, we acknowledge support by the U.S. Department of Energy Basic Energy Office of Science, Basic Energy Sciences (Grant No. DEFG02-00-ER45841) and, for sample synthesis, NSF Grants No. ECCS 1906253 and the Gordon and Betty Moore Foundation’s EPiQS Initiative (Grant No. GBMF9615 to L.N.P.). This research is funded in part by QuantEmX Travel Grants from the Institute for Complex Adaptive Matter. A portion of this work was performed at the National High Magnetic Field Laboratory (NHMFL), which is supported by National Science Foundation Cooperative Agreement No. DMR-1644779 and the state of Florida. We thank S. Hannahs, T. Murphy, A. Bangura, G. Jones, and E. Green at NHMFL for technical support. We also thank J. K. Jain for illuminating discussions.
\end{acknowledgments}

%\bibliography{ref}% Produces the bibliography via BibTeX.

\end{document}

\setcounter{page}{1}

\title[]{Supplemental Material for ``Experiments on Delocalization and Universality of the Fractional Quantum Hall Plateau-to-Plateau Transitions"}
\author{P. T. \surname{Madathil}}
\author{K. A. \surname{Villegas Rosales}}
\author{C. T. \surname{Tai}}
\author{Y. J. \surname{Chung}}
\author{L. N. \surname{Pfeiffer}}
\author{K. W. \surname{West}}
\author{K. W. \surname{Baldwin}}
\author{M. \surname{Shayegan}}
\affiliation{Department of Electrical Engineering, Princeton University, Princeton, New Jersey 08544, USA}
\date{\today}

\maketitle  

\noindent\begin{picture}(0,0)
\put(0,-390){\begin{minipage}{\textwidth}
\centering
\includegraphics[width=1\textwidth]{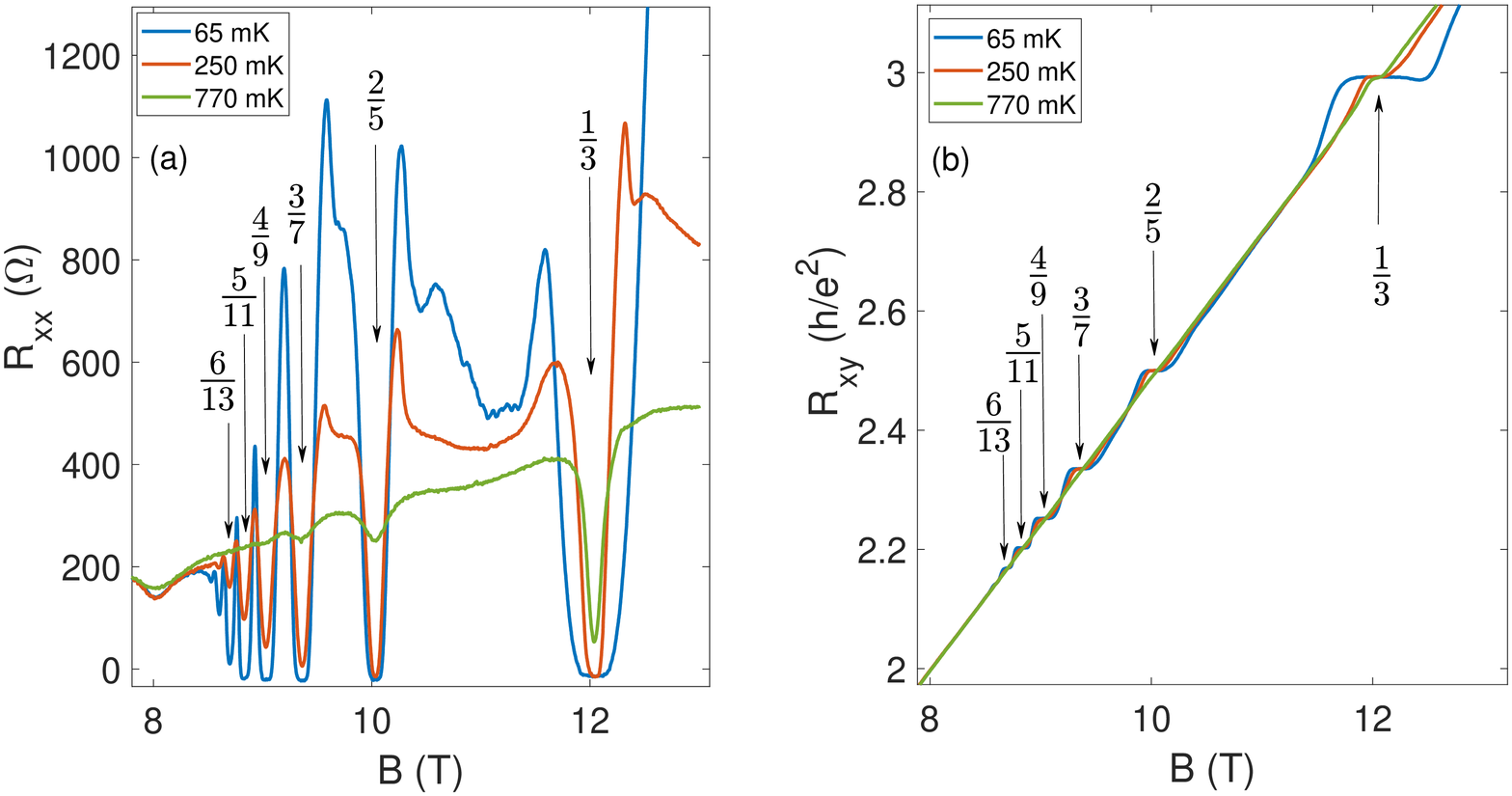}
\captionof{figure}{(a) Longitudinal resistance ($R_{xx}$) vs. magnetic field ($B$) for a 2DES confined to a 50-nm-wide GaAs QW at three different temperatures, $T\simeq$ 65, 250, and 770 mK. (b) Hall resistance ($R_{xy}$) vs. $B$ data at the same temperatures. }
\end{minipage}}
\end{picture}%
%\section{Data for the 50-nm-wide quantum well}

%\begin{figure*}[t]
%\centering
%\includegraphics[width=1\textwidth]{rxx_rxy_trace.eps}
%\centering

%  \caption{\label{fig:wide} 
%(a) Hall resistance ($R_{xy}$) vs. magnetic field ($B$) for a 2DES confined to a 50-nm-wide GaAs QW at three different temperatures, $T\simeq$ 65 mK,$T\simeq$ 250 mK, and $T\simeq$ 770 mK.(b) Longitudinal resistance ($R_{xx}$) vs. magnetic field ($B$) data for the same temperatures. 
 % }
%  \label{fig:Ip}
%\end{figure*}

\begin{minipage}[t]{\textwidth}

\section{Extraction of $\kappa$ from Hall data for the 50-nm-wide quantum well}
\setlength{\parindent}{5mm}
\setlength{\parskip}{5mm}

In the main text, we extracted the exponent ($\kappa$) from the temperature dependence of the inverse of the half-width of longitudinal magneto-resistance ($R_{xx}$) between two successive fractional quantum Hall states (FQHSs). $\kappa$ can be extracted from the temperature dependence of the Hall resistance ($R_{xy}$) as well. The derivative of $R_{xy}$ with respect to the magnetic field $B$ exhibits a maximum at the critical magnetic field ($B_c$), and the value of this maximum should show a power-law divergence with temperature with the same critical exponent as $R_{xx}$ data, i.e. $\frac{dR_{xy}}{dB}|_{B = B_c} \propto T^{-\kappa}$ \cite{pruisken1988universal,huckestein1995scaling,wei1988experiments,li2005scaling}. Here in this Supplemental Material, we describe the procedure for extracting $\kappa$ from the $R_{xy}$ data and present the values of $\kappa$ for the 50-nm-wide quantum well.

Figures S1(a,b) show $R_{xx}$ and $R_{xy}$ vs. $B$ traces for our two-dimensional electron system (2DES) confined to a 50-nm-wide, GaAs quantum well. Data are shown at temperatures $T\simeq$ 65, 250, and 770 mK, represented by the blue, red and green traces, respectively. The $R_{xx}$ data reveal the presence of intermediate features between $\nu = 1/3$ and 2/5, and 2/5 and 3/7. This is also reflected in $R_{xy}$ where the plateau-to-plateau transitions between these fillings are disrupted by the presence of the classical Hall slope even at the lowest temperatures. In contrast, the transition between FQHSs of intermediate strength, e.g. between $\nu = 4/9$ and 5/11, is step-like, indicative of an ideal plateau-to-plateau transition.  We discuss these features in detail in Figs. S2 and S3, respectively.
\end{minipage}
\clearpage

Figure S2 includes the $R_{xy}$ data between $\nu = 1/3$ and 2/5. The blue trace is $R_{xy}$ vs. $B$ and the red trace is $dR_{xy}/dB$ vs. $B$. In the derivative of $R_{xy}$, we see two local maxima, one closer to 2/5 at $B$ = 10.32 T and another closer to 1/3 at $B$ = 11.57 T. We also observe a broad, nearly constant value of $dR_{xy}/dB \simeq$ 6.23 k$\Omega$/T  between $B \simeq 10.5$ T and $B \simeq 11.3$ T, corresponding to the classical Hall slope. There are also small dips at $B$ = 11.04 T and 10.44 T corresponding to the fractions 4/11 and 5/13. The transition between 1/3 and 2/5 is mediated by a classical Hall line and the conventional plateau-to-plateau transition is split into a plateau-to-classical and classical-to-plateau transition. This clearly indicates that in ultra-high-quality samples, the transitions between strong FQHSs, such as the transition between $\nu = 1/3$ and 2/5 is influenced by the presence of these additional states. We then proceed to do a temperature dependence of $\frac{dR_{xy}}{dB}|_{B = 10.32\textnormal{T}}$ and $\frac{dR_{xy}}{dB}|_{B = 11.57\textnormal{T}}$ to extract exponents ($\kappa$) for both maxima and report the values in Fig. S4(b). 

\begin{figure}[h]
\centering
\includegraphics[width=0.5\textwidth]{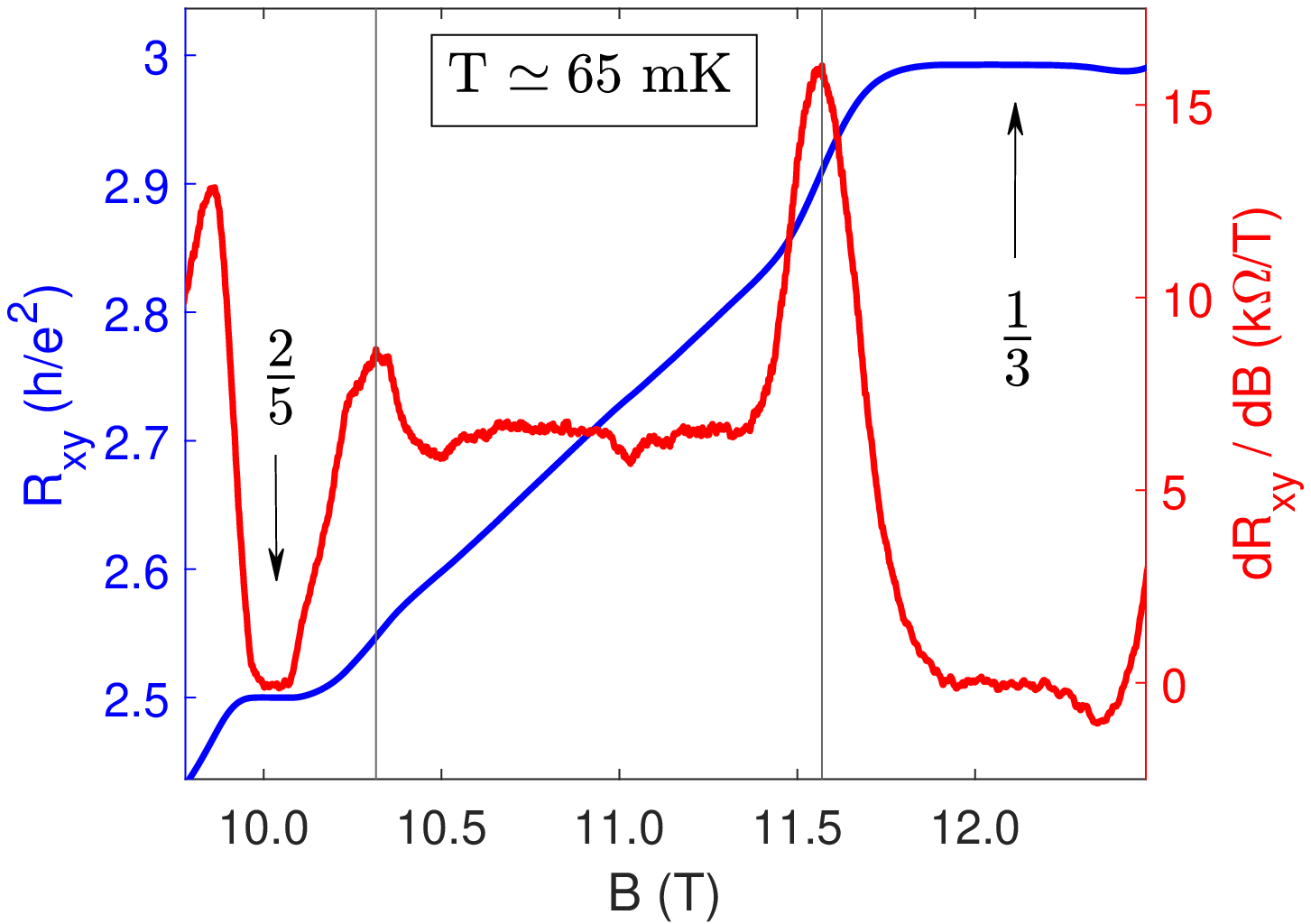}
\centering
  \caption{\label{Ip} 
 The blue trace is the smoothed $R_{xy}$ between $\nu = 1/3$ and 2/5. The red trace is the corresponding $dR_{xy}/dB$ vs. $B$. The vertical lines mark $B$ at which $dR_{xy}/dB$ has a maximum, at $B$ = 10.32 T and $B$ = 11.57 T. 
  }

  \label{fig:Ip}
\end{figure}

Figure S3 shows the $R_{xy}$ data for the transition between $\nu = 4/9$ and 5/11. The blue trace is $R_{xy}$ vs. $B$ and the red trace is $dR_{xy}/dB$ vs. $B$. This transition, between FQHSs of intermediate strength, is a more ideal plateau-to-plateau transition, with only one $dR_{xy}/dB$ maximum corresponding to the critical field, $B_c = 8.92$ T. The exponent $\kappa$ extracted for this transition ($\kappa\simeq 0.44$) is indeed closer to the theoretically predicted value.

\begin{figure}[h]
\centering
\includegraphics[width=0.5\textwidth]{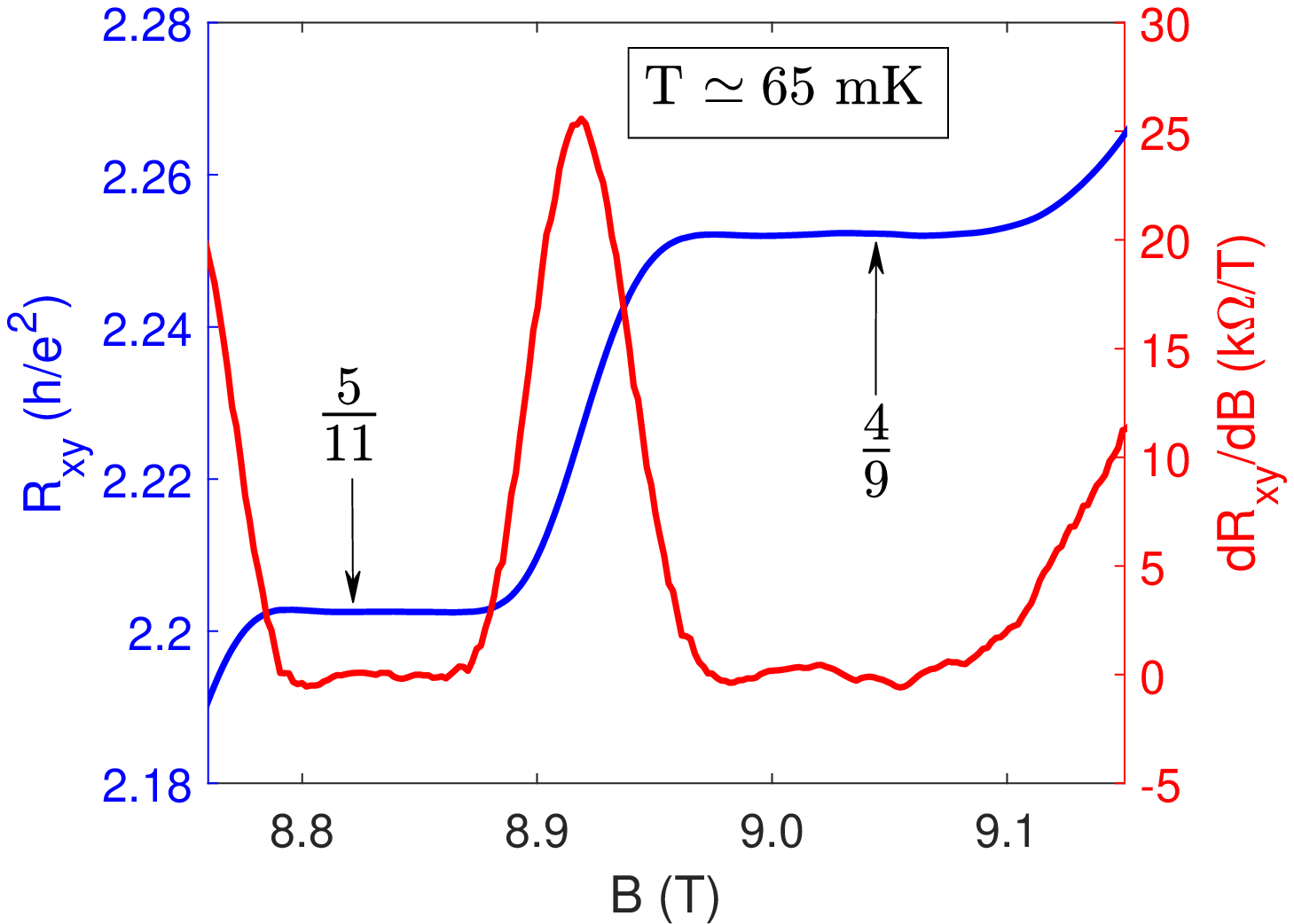}
\centering
  \caption{\label{Ip} 
The blue trace is the smoothed $R_{xy}$ between $\nu = 4/9$ and 5/11, and the red trace is the corresponding $dR_{xy}/dB$ vs. $B$. Here, there is only one $dR_{xy}/dB$ maximum at $B$ = 8.92 T.  
  }
  \label{fig:Ip}
\end{figure}

\begin{figure*}[t]
\centering
\includegraphics[width=1\textwidth]{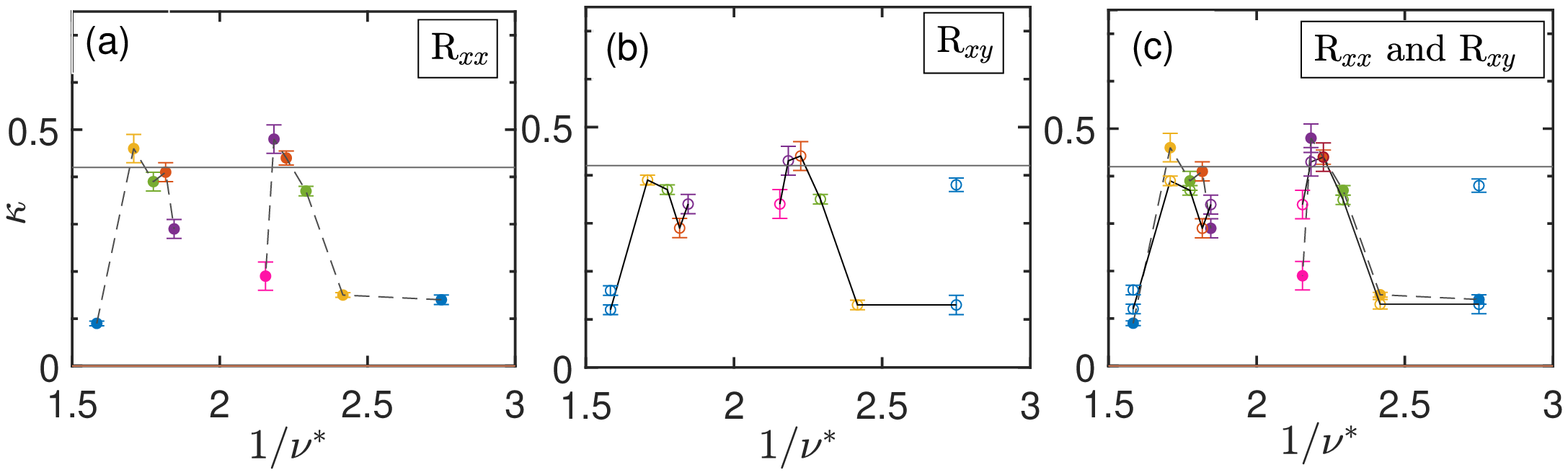}
\centering
  \caption{\label{Ip} 
(a) The extracted $\kappa$ from $R_{xx}$ data, according to the relation $1/\Delta \propto T^{-\kappa}$, for the 50-nm-wide quantum well are plotted vs. $1/\nu^*$ defined as $1/\nu^* = (1/\nu_1 + 1/\nu_2)/2$, where $\nu_1$ and $\nu_2$ are the fillings of two consecutive FQHSs. This is the same figure as Fig. 2(c) of the main text, and the colors of data points for $2<1/\nu^*<3$ represent the colors of data presented in Fig. 1(d) of the main text for different transitions. The dashed lines connecting the data points are guides to the eye. The black, horizontal line at $\kappa = 0.42$ represents the expected exponent. (b) The extracted $\kappa$ from $R_{xy}$ data, according to the relation $\frac{dR_{xy}}{dB}|_{B = B_c} \propto T^{-\kappa}$, are plotted vs. $1/\nu^*$.  (c) Summary and comparison of  the extracted $\kappa$ vs. 1/$\nu^*$ from the temperature dependence of $R_{xx}$, represented by closed symbols, and $R_{xy}$, represented by open symbols.
  }
  \label{fig:Ip}
\end{figure*}

Figure S4(a) summarizes the values of $\kappa$ obtained from $R_{xx}$, according to the relation $1/\Delta \propto T^{-\kappa}$,  for the 50-nm-wide quantum well. The x-axis is defined as $1/\nu^* = (1/\nu_1 + 1/\nu_2)/2$, where $\nu_1$ and $\nu_2$ are the fillings of two consecutive FQHSs and the y-axis gives the $\kappa$ values. The dashed lines connecting the data points are guides to the eye and the horizontal line at $\kappa = 0.42$ represents the expected exponent from theoretical calculations. In Fig. S4(b), we present the exponents obtained from $R_{xy}$ data, according to the relation, $\frac{dR_{xy}}{dB}|_{B = B_c} \propto T^{-\kappa}$. It is evident from Figs. S4(a) and S4(b) that $\kappa$ values extracted from $R_{xy}$ qualitatively follow the same trend, and are quantitatively close to the values extracted from the temperature dependence of $R_{xx}$.  This is clearly evident from Fig. S4(c) which presents $\kappa$ extracted from $R_{xy}$ and $R_{xx}$ data in the same plot.

It is noteworthy that, for the transition between the strongest FQHSs, namely the 1/3 to 2/5  transition, the $R_{xy}$ data yield two different values of $\kappa$ for the two local maxima in $dR_{xy}/dB$ seen in Fig. S2. For the $dR_{xy}/dB$ maximum at $B=11.75$ T, which is closer to the stronger fraction ($\nu=1/3$), we find $\kappa\simeq 0.38$, while the $dR_{xy}/dB$ maximum at $B=10.32$ T, which is closer to the weaker fraction ($\nu=2/5$), yields $\kappa \simeq 0.13$. The reason for this difference is unclear. We note that the larger $\kappa \simeq 0.38$ is closer to the theoretically-expected value. However, for the $\nu=2/3$ to 3/5 transition, where we also see two $dR_{xy}/dB$ maxima (data not shown), the two extracted $\kappa$ values, 0.16 for the maximum near 2/3, and 0.12 for the maximum near 3/5, are both much smaller than the expected $\kappa =0.42$, and close to the $\kappa$ value extracted from $R_{xx}$; see the data points at $1/\nu^* \simeq 1.58$ in Fig. S4(c). These observations highlight the complexity of the transitions between strong FQHSs, brought about by the presence of many-body induced states in the transition regions.